\documentclass{emulateapj}
\usepackage{color}
\usepackage{times}
\usepackage{fancyhdr}
\usepackage{float}
\usepackage{alltt}
\usepackage{listings}
\usepackage{enumerate}
\usepackage[T1]{fontenc}
\usepackage{ae}
\usepackage{rotating}
\usepackage{subfigure}
\usepackage{amsmath}
\usepackage{amssymb}
\usepackage{graphicx}
\usepackage{dcolumn}
\usepackage{bm}
\usepackage{aas_makros_pk}
\usepackage{hyperref}
\usepackage{natbib}

\def\be{\begin{equation}}
\def\ee{\end{equation}}
\def\beg{\begin{align}}
\def\eeg{\end{align}}
\def\bi{\begin{itemize}}
\def\ei{\end{itemize}}
\def\ben{\begin{enumerate}[1.]}
\def\een{\end{enumerate}}

\newcommand{\bo}{\raise-1mm\hbox{\Large$\Box$}}
\newcommand{\sci}[2]{#1 \times 10^{#2}}
\newcommand{\rthz}{\mathrm{Hz}^{-\frac{1}{2}}}
\newcommand{\hrss}{h_{\mathrm{rss}}}

\newcommand{\hrssn}{h_{\mathrm{rss}}^{90\%}}

\newcommand{\egwn}{E_{\mathrm{GW}}^{90\%}}
\newcommand{\egw}{E_{\mathrm{GW}}}
\newcommand{\eem}{E_{\mathrm{EM}}}

\newcommand\ligodoc{P0900192}

\begin{document}

\title{Search for Gravitational Wave Bursts from Six Magnetars}

\author{J.~Abadie\altaffilmark{1}, 
B.~P.~Abbott\altaffilmark{1}, 
R.~Abbott\altaffilmark{1}, 
M.~Abernathy\altaffilmark{2}, 
T.~Accadia\altaffilmark{3}, 
F.~Acernese\altaffilmark{4ac}, 
C.~Adams\altaffilmark{5}, 
R.~Adhikari\altaffilmark{1}, 
C.~Affeldt\altaffilmark{6,7}, 
B.~Allen\altaffilmark{6,8,7}
% ...
}

\altaffiltext{1}{LIGO - California Institute of Technology, Pasadena, CA  91125, USA }
\altaffiltext{2}{University of Glasgow, Glasgow, G12 8QQ, United Kingdom }
\altaffiltext{3}{Laboratoire d'Annecy-le-Vieux de Physique des Particules (LAPP), Universit\'e de Savoie, CNRS/IN2P3, F-74941 Annecy-Le-Vieux, France}
\altaffiltext{4}{INFN, Sezione di Napoli $^a$; Universit\`a di Napoli 'Federico II'$^b$ Complesso Universitario di Monte S.Angelo, I-80126 Napoli; Universit\`a di Salerno, Fisciano, I-84084 Salerno$^c$, Italy}
\altaffiltext{5}{LIGO - Livingston Observatory, Livingston, LA  70754, USA }
\altaffiltext{6}{Albert-Einstein-Institut, Max-Planck-Institut f\"ur Gravitationsphysik, D-30167 Hannover, Germany}
\altaffiltext{7}{Leibniz Universit\"at Hannover, D-30167 Hannover, Germany }
\altaffiltext{8}{University of Wisconsin--Milwaukee, Milwaukee, WI  53201, USA }
% ...
\affil{The LIGO Scientific Collaboration and the Virgo Collaboration}

\author{R. L.~Aptekar\altaffilmark{81}}
\altaffiltext{81}{Ioffe Physico-Technical Institute, Russian Academy of Science, St.Petersburg, 194021, Russia}
 
\author{W. V.~Boynton\altaffilmark{82}}
\altaffiltext{82}{Department of Planetary Sciences, University of Arizona, Tucson, AZ 85721, USA}
  
\author{M. S.~Briggs\altaffilmark{83}}
\altaffiltext{83}{CSPAR, University of Alabama in Huntsville, Huntsville, Alabama, USA}

\author{T. L.~Cline\altaffilmark{43}}
  
\author{V.~Connaughton\altaffilmark{83}}   
  
\author{D. D.~Frederiks\altaffilmark{81}}  

\author{N.~Gehrels\altaffilmark{84}}
\altaffiltext{84}{NASA-GSFC, Code 661, Greenbelt, MD 20771, USA}  

\author{J. O~Goldsten\altaffilmark{85}} 
\altaffiltext{85}{The Johns Hopkins University Applied Physics Laboratory, Laurel, MD 20723, USA}

\author{D.~Golovin\altaffilmark{86}} 
\altaffiltext{86}{Institute for Space Research, Profsojuznaja 84/32 117997 Moscow, Russia}
 
\author{A.J.~{van der Horst}\altaffilmark{87}}
\altaffiltext{87}{NASA Postdoctoral Program Fellow, NASA Marshall Space Flight Center, Huntsville, AL 35805, USA}
 
\author{K. C.~Hurley\altaffilmark{88}}
\altaffiltext{88}{University of California-Berkeley, Space Sciences Lab, 7 Gauss Way, Berkeley, CA 94720, USA}

\author{Y.~Kaneko\altaffilmark{89}}
\altaffiltext{89}{Sabanc\i University, Orhanl\i-Tuzla 34956 \.Istanbul, Turkey}

\author{A.~{von Kienlin}\altaffilmark{90}}
\altaffiltext{90}{Max-Planck Institut f\"ur extraterrestrische Physik, Giessenbachstrasse 1, 85748 Garching, Germany}

\author{C.~Kouveliotou\altaffilmark{91}}
\altaffiltext{91}{Space Science Office, VP62, NASA Marshall Space Flight Center, Huntsville, AL 35812, USA}

\author{H. A.~Krimm\altaffilmark{92}} 
\altaffiltext{92}{CRESST and NASA Goddard Space Flight Center, Greenbelt, MD 20771, USA}

\author{L.~Lin\altaffilmark{83,93}} 
\altaffiltext{93}{The National Astronomical Observatories, Chinese Academy of Sciences, Beijing 100012, China}

\author{I.~Mitrofanov\altaffilmark{85}}

\author{M.~Ohno\altaffilmark{94}}
\altaffiltext{94}{Institute of Space and Astronautical Science, Japan Aerospace Exploration Agency, 3-1-1 Yoshinodai, Chuo-ku Sagamihara, Kanagawa 252-5120, Japan}

\author{V. D.~Pal'shin\altaffilmark{81}}

\author{A.~Rau\altaffilmark{90}}

\author{A.~Sanin\altaffilmark{85}}

\author{M. S.~Tashiro\altaffilmark{95}}
\altaffiltext{95}{Department of Physics, Saitama University, 255 Shimo-Okubo, Sakura, Saitama, 338-8570, Japan}

\author{Y.~Terada\altaffilmark{95}}

\author{K.~Yamaoka\altaffilmark{96}}
\altaffiltext{96}{Department of Physics and Mathematics, Aoyama Gakuin University, 5-10-1 Fuchinobe, Chuo-ku, Sagamihara, Kanagawa 252-5258, Japan}

\keywords{gravitational waves --- stars: magnetars}
\shorttitle{Search for Gravitational Wave Bursts from Six Magnetars}
\shortauthors{The LIGO Scientific Collaboration and the Virgo Collaboration}

\begin{abstract}
Soft gamma repeaters (SGRs) and anomalous X-ray pulsars (AXPs) are thought to 
be magnetars: neutron stars powered by extreme magnetic fields.  These rare objects are 
characterized by repeated and sometimes spectacular gamma-ray bursts.  The 
burst mechanism might involve crustal fractures and excitation of non-radial modes 
which would emit gravitational waves (GWs).  We present the results of a search for 
GW bursts from six galactic magnetars that is sensitive to neutron star $f$-modes, 
thought to be the most efficient GW emitting oscillatory modes in compact stars.  
One of them, SGR 0501+4516, is likely $\sim1$\,kpc from Earth, an order of 
magnitude closer than magnetars targeted in previous GW searches.    A second, 
AXP 1E 1547.0$-$5408, gave a burst with an estimated isotropic energy $>10^{44}$
\,erg which is comparable to the giant flares.  We find no evidence of GWs associated 
with a sample of 1279 electromagnetic triggers from six magnetars occurring 
between November 2006 and June 2009, in GW data from the LIGO, Virgo, and 
GEO600 detectors.  Our lowest model-dependent GW emission energy upper limits 
for band- and time-limited white noise bursts in the detector sensitive band, and for $f
$-mode ringdowns (at 1090\,Hz), are $\sci{3.0}{44} d_\mathrm{1}^2 $\,erg and $
\sci{1.4}{47} d_\mathrm{1}^2 $\,erg respectively, where $d_\mathrm{1} = 
\frac{d_{\mathrm{0501}}}{1\,\mathrm{kpc}}$ and $d_{\mathrm{0501}}$ is the distance 
to SGR 0501+4516.  These limits on GW emission from $f$-modes are an order of 
magnitude lower than any previous, and approach the range of electromagnetic 
energies seen in SGR giant flares for the first time.
\end{abstract}

\section{Introduction}

Magnetars are isolated neutron stars (NS) powered by extreme magnetic fields ($\sim
\nolinebreak10^{15}$\,G)\,\citep{duncan92}.  The magnetar model explains the 
observed properties of two classes of rare objects, the soft gamma repeaters (SGRs) 
and the anomalous X-ray pulsars (AXPs): compact X-ray sources with long rotation 
periods and rapid spindowns which sporadically 
emit short ($\approx0.1$\,s) bursts of soft gamma rays (for a review see\,
\citealt{mereghetti08}.)  Fewer than twenty SGRs and AXPs are known.  The total isotropic
burst energies rarely exceed  $10^{42}$\,erg. 
However, three extraordinary ``giant flares'' (GFs) have been observed in $\sim$30 years from 
SGRs in our galaxy and the Large Magellanic Cloud:  one from SGR 0526$-$66 in 1979 
with an observed total isotropic energy of $\sim\sci{1.2}{44}d_\mathrm{55}^2$\,erg\,\citep{mazets79}; one from 
SGR 1900+14 in 1998 with $\sci{4.3}{44}d_\mathrm{15}^2$\,erg\,\citep{tanaka07}; and a 
spectacular one from SGR 1806$-$20 in 2004 with $\sim\sci{5}{46}d_\mathrm{15}^2$\,erg\,
\citep{terasawa05} where $d_n = d/({n\,\mathrm{kpc}})$.  There is also evidence that some short gamma ray bursts (GRBs) 
were in fact extragalactic GFs.  GRB 070201 might have been a GF located in 
the Andromeda galaxy with an isotropic energy of $\sci{1.5}{45}$\,erg\,\citep{mazets08, 
S5GRB070201};  and GRB 051103 might have been a GF in M81 with an energy of 
$\sci{7.5}{46}$\,erg\,\citep{frederiks07b}.

\begin{figure}[!t]
\begin{center}
\includegraphics[angle=0,width=95mm,  clip=true, viewport=.5in 2.5in 8.5in 8.5in]{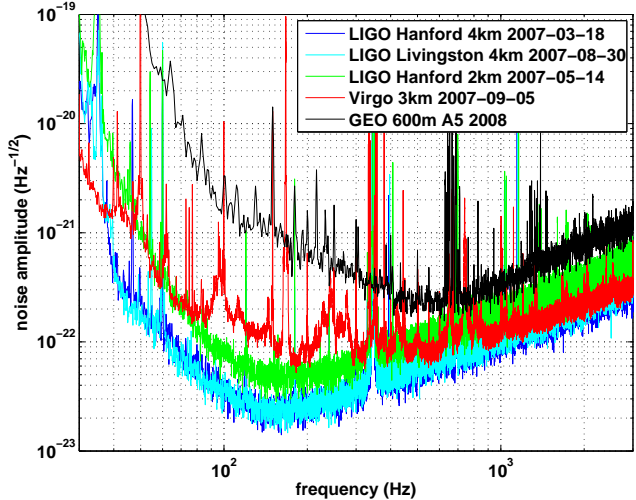}
\caption{ Best detector noise spectra from the LIGO and Virgo detectors 
during S5/VSR1 and the GEO600 detector during A5.  } \label{fig:noise}
\end{center}
\end{figure}

Although still poorly understood, magnetars are promising candidates for the first direct 
gravitational wave (GW) detection for several reasons.   First, a sudden localized energy 
release could excite non-radial pulsational NS modes. Bursts may be caused by 
untwisting of the global interior magnetic field and associated cracking of the solid NS 
crust\,\citep{thompson95}, or global reconfiguration of the internal magnetic field and 
associated deformation of the NS hydrostatic equilibrium\,\citep{ioka01, corsi11}.  The lowest-
order GW emitting mode, the $f$-mode, is damped principally via GW emission and 
would ring down with a predicted damping time of 100--400\,ms and with a frequency in 
the 1--3\,kHz range depending on the nuclear equation of state and NS composition\,
\citep{benhar04}, putting these signals in the band of interferometric GW 
detectors (see Figure\,\ref{fig:noise}).  Second, precise sky locations and trigger times 
from electromagnetic (EM) bursts allow us to reduce the false-alarm rate and increase sensitivity relative to 
all-sky all-time searches such as\,\cite{S5y2Burst}. Finally, magnetars are among the 
closest of potential GW burst sources.  

GW signals from magnetars would give us a new window through
which to probe the stellar physics and structure.  However,
quantitative predictions or constraints on the amplitude of GW emission 
associated with magnetar bursts are relatively few and highly uncertain (see e.g.\,\citealt{ioka01, owen05, 
horowitz09, corsi11, kashiyama11, levin11} );
hence it is not clear when we might begin to expect a detection.

It may turn out that the magnetar burst mechanism does not excite global NS
$f$-modes.  If the outburst dynamics are confined to surface layer modes,
the crust torsional oscillations might emit GWs at frequencies of
$\sim10-2000$\,Hz\,\citep{mcdermott88}.   It is also possible that
although the crust is a plausible site for triggering, bursts are confined to the 
magnetosphere\citep{Lyutikov:2005un}, although even in this case $f$-modes 
might be excited either directly or via crust/core hydromagnetic coupling.
Finally we note that it is not yet clear if GFs and common 
bursts are caused by the same mechanism.  The lack of theoretical 
understanding underlines the importance of observational 
constraints on GW emission.

We present results from a search for GW bursts associated with magnetar EM
bursts using data from the second year of LIGO's fifth science run (S5y2,
\citealt{ligoDetector}); Virgo's first science run (VSR1, \citealt{virgo08});
and the subsequent LIGO and GEO \textit{astrowatch} period (A5), during which
the principal goal was detector commissioning, not data collection.   The S5y2 epoch 
involved the three LIGO detectors:  a 4\,km interferometer in Louisiana and two 
interferometers (4\,km and 2\,km) in Washington.  The VSR1 epoch added the Virgo 
3\,km detector to the global network.  The A5 epoch included only the LIGO 2\,km 
detector and the GEO 600\,m detector\,\citep{grote10}.  The Virgo and GEO600 detectors 
are located in Italy and Germany, respectively.  

This is the third search for GWs from magnetars sensitive to $f$-mode
ringdowns.  The first\,\citep{s5y1sgr} included the 2004 SGR
1806$-20$ GF, a 2006 storm of bursts from SGR 1900+14, and 188 other
events from SGRs 1806$-$20 and 1900+14 occurring before November 2006.  Upper
limits on $f$-mode GW energy emission at 1090\,Hz ranged from
$\sci{2.4}{48}$\,erg to $\sci{2.6}{51}$\,erg, and upper limits on band- and
time-limited white noise bursts at 100--200\,Hz ranged from
$\sci{3.1}{45}$\,erg to $\sci{7.3}{47}$\,erg.  The second\,\citep{stackSgr} focused on the 2006 SGR 1900+14 storm,
``stacking'' GW data corresponding to individual bursts in the storm's EM
lightcurve\,\citep{stackMethod}. An upper limit on $f$-mode emission at 1090\,Hz of $\sci{1.2}{48}$\,erg per burst was set
on a stack of the 11 brightest storm bursts, an order 
of magnitude lower than the unstacked limit on the storm.

\begin{figure}[!t]
\begin{center}
\includegraphics[angle=0,width=95mm, clip=true, viewport=0.5in 0.3in 8.5in 6in]{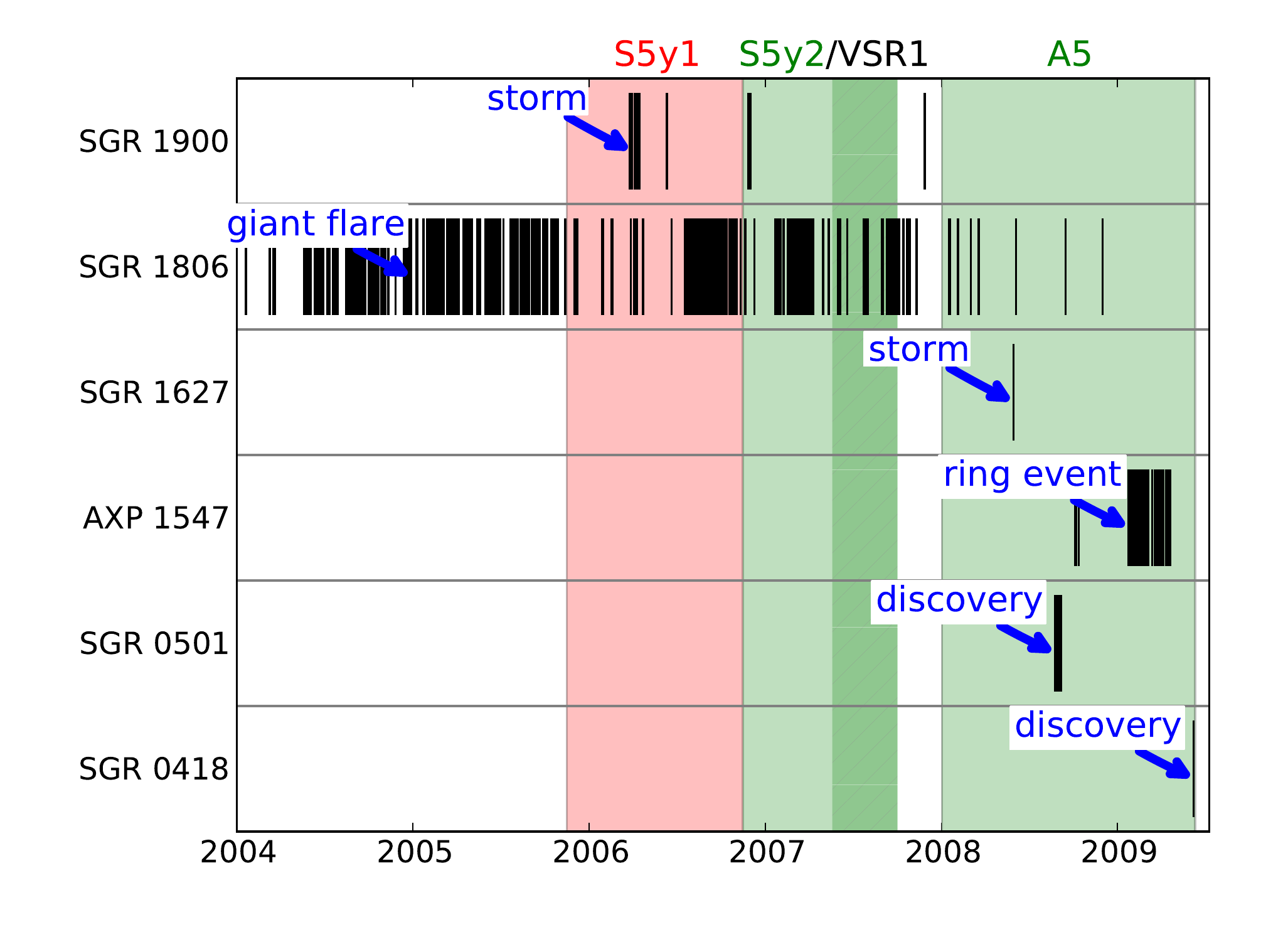}
\caption{ Each mark represents a burst from one of the six magnetar sources.  Exceptional 
events are annotated in the figure; SGR 0501+4516 and SGR 0481+5279 were discovered in the A5 epoch.  
The LIGO S5y1 epoch was the subject of the first $f$-
mode search\,\citep{s5y1sgr}.  The current search includes bursts which occurred during the 
LIGO S5y2 and Virgo VSR1 epochs for which usable data were available, as well as the A5 
astrowatch commissioning period.  The VSR1 epoch, which is a subset of the S5y2 epoch, is 
indicated by cross-hatching and darker shading. (NB: unabbreviated source names are 
given in Table\,\ref{table:magnetars}.)  } \label{fig:activity}
\end{center}
\end{figure}

During the S5y2, VSR1 and A5 epochs of the search we present here, 1217 soft gamma-ray 
bursts from six magnetars were listed by the interplanetary network of satellites\,
\footnote{http://ssl.berkeley.edu/ipn3} or IPN (Table\,\ref{table:magnetars} and Figure\,\ref{fig:activity}).  Four of 
the sources are being examined for GW signals 
for the first time.  Two of those (SGR 0501+4516 and SGR 
0418+5729) are thought to be much closer to Earth than
SGRs examined in previous GW searches. 
SGR 0501+4516 might be associated with the supernova remnant HB9\,\citep{gcn8149}, which is $(800\pm 400)$\,pc from Earth\,
\citep{leahy07}; proper motion measurements could exclude this association.  
The probable locations of both SGR 0501+4516 and SGR 0418+5729 in the Perseus arm of our galaxy 
imply distances of $\sim$1--2\,kpc\,\citep{vanDerHorst10}.  
AXP 1E 1547.0$-$5408 (also known as SGR 1550--5418) 
gave two exceptional bursts on 2009 January 22.  Observations of expanding 
rings around the source, caused by X-ray scattering off dust 
sheets, set the source distance at 4--5\,kpc and imply an EM energy for one or both of these 
``ring event'' bursts of $10^{44-45}$\,erg\,\citep{tiengo10}, comparable to the GFs.  
In addition to the IPN triggers, we include eight 
triggers from the Fermi GBM detector: 
seven bright AXP 1E 1547.0$-$5408 bursts and one SGR 0418+5729 burst.  
We also identified 54 individual peaks in 
a storm from SGR 1627$-$41 lasting $\sim2000$\,s by combining the 15--25\,keV and 25--50\,keV 
Swift/BAT 64\,ms-binned light curves\footnote{http://heasarc.gsfc.nasa.gov/FTP/swift/data/
obs/2008\_05/00312582000} \footnote{http://heasarc.gsfc.nasa.gov/FTP/swift/data/obs/
2008\_05/00090056009} and selecting peaks above 450 counts / 64 ms.  The search thus 
includes a grand total of 1279 EM triggers. 

\begin{table*}
\begin{minipage}{\textwidth}
\begin{center}
\caption[Summary of SGR properties]{ Summary of source sky locations and estimated 
distances.  The nominal distances $d_N$ are the distance used in the search for setting 
upper limits.  Energy upper limits can be scaled to any distance $d$ via the factor $d^2 / 
d_N^2$. Some EM triggers occurred when there was no GW data available (i.e. $N=0$).}
\begin{tabular}{ll|cc|c|ccc}
 \hline \hline
 \textbf{Source} & \textbf{Position}  & \multicolumn{2}{c}{\textbf{Distances (kpc)}}   & \textbf{EM triggers} & \multicolumn{3}{c}{\textbf{Analyzed with $N$ detectors}}  \\

 &   J2000  &     \textbf{Estimated} & \textbf{Nominal}  & \textbf{total} & \textbf{$N=1$} & \textbf{$N=2$} & \textbf{$N\ge 3$}     \\

 \hline
SGR 0418+5729\footnote{position:  \cite{atel2159}; distance: \cite{esposito10, vanDerHorst10}} & $04^{\mathrm{h}} 18^{\mathrm{m}} 33.867\pm0.35^{\mathrm{s}}$ & $\sim$2  & 2  & 3 & 3 & - & -    \\
             & $+57^\circ 32' 22.91\pm0.35''$ &   &     & & & & \\
             \hline 
              
SGR 0501+4516\footnote{position: \cite{gcn8148}; distance: \cite{vanDerHorst10, gcn8149, leahy07}} & $05^{\mathrm{h}} 01^{\mathrm{m}} 06.8\pm1.4^{\mathrm{s}}$ & $\sim$2, 0.8$\pm$0.4  & 1  & 166 & 105 & 24 & -  \\ 
             & $+45^\circ 16' 35.4\pm1.4''$ &   &    & & & & \\
             \hline   
             
AXP 1E 1547.0$-$5408\footnote{position: \cite{camilo07}; distance: \cite{tiengo10, camilo07, gelfand07}} & $15^{\mathrm{h}} 50^{\mathrm{m}} 54.11\pm0.01^{\mathrm{s}}$ & 4-5, 9, 4  & 4  & 844 & 315 & 512 & -   \\
             & $-54^\circ 18' 23.7\pm0.1''$ &    &   & & & &\\
             \hline        
             
SGR 1627--41\footnote{position: \cite{wachter04}; distance: \cite{corbel99}} & $16^{\mathrm{h}} 35^{\mathrm{m}} 51.84\pm0.2^{\mathrm{s}}$ & 11$\pm$0.3  & 11  & 56 & - & 56 & -   \\
             & $-47^\circ 35' 23.31\pm0.2''$ &   &    & & & & \\
             \hline
             
SGR 1806$-$20\footnote{position: \cite{kaplan02a}; distance: \cite{bibby08, cameron05}} & $18^{\mathrm{h}} 08^{\mathrm{m}} 39.32\pm0.3^{\mathrm{s}}$ & 8.7$^{+1.8}_{-1.5}$, 6.4--9.8   & 10  & 207  & 11 & 36 & 136   \\
             & $-20^\circ 24' 39.5\pm0.3''$ &   &    & & & & \\
             \hline
             
SGR 1900+14\footnote{position: \cite{frail99}; distance: \cite{marsden01b, vrba00}}  & $19^{\mathrm{h}} 07^{\mathrm{m}} 14.33\pm0.15^{\mathrm{s}}$
& 3--9, 12--15  & 10    & 3 & - & 1 & - \\
             & $+09^\circ 19' 20.1\pm0.15''$ &  &      & & & & \\     
             \hline

\end{tabular}
\label{table:magnetars}
\end{center}
\end{minipage}
\end{table*}

\section{Method} \label{sec:method}

We analyze magnetar bursts using the strategy from\,\cite{s5y1sgr}, 
which is less dependent on a particular emission model than the stacking 
approach of\,\cite{stackSgr}.  The analysis is performed by the Flare 
pipeline\,\citep{kalmus07, kalmus08}, which produces a time-frequency excess power
pixel map from calibrated 
detector data streams in the Fourier basis.   Pixels are 
characterized by excess power relative to the background (``loudness'') and 
loud adjacent pixels are grouped into ``events.''  The generalized pipeline 
accepts arbitrary networks of GW detectors by including detector noise floor 
measurements and antenna responses in the detection statistic\,\citep{flaredetstat}. 
We divide the search into three frequency bands:  1--3\,kHz 
where $f$-modes are predicted to ring; and 100--200\,Hz and 100--1000\,Hz.  
We include the latter two frequency bands in order to search also at lower 
frequencies where the detectors are most sensitive (see Figure\,\ref{fig:noise}).  
Although there are no predictions of GW \textit{burst} signals from magnetars 
at these lower frequencies, we note that quasiperiodic oscillations (QPOs), 
lasting for tens of seconds and possibly associated with stellar torsional 
modes, have been observed in GF EM tails at frequencies as low as 
18\,Hz and as high as 1800\,Hz\,\citep{strohmayer05, israel05, steiner09}.  
QPOs in the tail of the 
2004 GF were targeted by a tailored GW search\citep{qpo} distinct from 
the one presented here.

As in\,\cite{s5y1sgr}, we choose 4\,s signal regions centered on each EM
trigger time.   
Delays between EM and GW emission are unlikely to be
significant\,\citep{stackMethod}; the 4\,s duration accounts for uncertainties in the geocentric
EM peak time due e.g. to satellite triggering algorithms and rounding.  Overlapping signal regions are
merged.  We analyze 1000\,s of
background on either side of each signal region (2000\,s total) in order to
estimate the significance of events in that signal
region. Background regions are not necessarily continuous,
as we require the same detector network coverage and data quality as for the
signal region; in addition, signal regions of other magnetar bursts are masked
out.  Signal and background regions are chosen after data quality cuts  have
been applied to the GW data, so as to remove data segments coincident with
instrumental or data acquisition problems, or excessive noise due to
challenging environmental conditions. For the S5y2 portion of this search, we
applied category 1 and 2 data quality cuts (i.e.\ cutting only data certain to
be unfit for analysis) as described in\,\cite{S5y2Burst}.  
For A5, which focused on detector commissioning, the boolean ``science mode'' 
designator and other basic data quality treatments were applied to the data, 
but the full categorical data quality treatment was not performed.   
Statistically significant events in the signal regions from any epoch are subject 
to follow-up investigations before being considered detection candidates. 
Follow-ups might include correlation with environmental data channels and
more refined estimates of significance.

We set model-dependent upper limits on $f$-mode ringdowns with 
circular and linear polarizations and frequencies sampling the range for 
$f$-modes (1--3\,kHz, which accounts for plausible NS equations of states 
and magnetic fields), and with a decay time constant of $\tau=200$\,ms.  We 
observed no more than 15\% degradation in strain upper limits 
using ringdowns with $\tau$ in the range 100--300\,ms as compared to the 
nominal value of 200\,ms.   We set additional limits on band- and time-
limited white noise bursts with 11\,ms and 100\,ms durations (motivated by 
observed rise times and durations of magnetar burst light curves) spanning 
the 100--200\,Hz and 100--1000\,Hz search bands.  While these frequencies are 
chosen principally to explore the detectors' most sensitive region below the 
$f$-mode frequencies, the observed range of QPO frequencies provides
astrophysical motivation.   Upper limits depend 
on the frequency sensitivity of the detectors (Figure \ref{fig:noise}).

Simulations are constructed using knowledge of the target magnetar's sky location and the 
EM burst time.  Following \cite{s5y1sgr}, $h_{\mathrm{rss}}^{2} = h_{\mathrm{rss+}}^{2} + 
h_{\mathrm{rss\times}}^{2}$, where $h_{\mathrm{rss+,\times}}^{2} = \int_{-\infty}^{\infty} 
h_{\mathrm{+,\times}}^{2} \mathrm{d}t$ and $h_{\mathrm{+,\times}}(t)$
are the two GW polarizations. The relationship between the GW polarizations and the 
detector response $h(t)$ to GW signals arriving from an altitude and azimuth
$(\theta,\phi)$ and with polarization angle $\psi$ is:
 \be
    h(t) = F^{+}(\theta, \phi, \psi) h_+(t)   +  F^{\times}(\theta, \phi, \psi)
 h_{\times}(t),
 \label{eq:hsim}
 \ee
where $F^{+}(\theta, \phi, \psi)$ and $F^{\times}(\theta, \phi,
\psi)$ are the antenna functions for the source at
$(\theta,\phi)$.   The polarization angle for each simulation
was randomly chosen from a flat distribution between 0 
and 2$\pi$.  The GW emission energy (if the integrand is averaged over inclination angle) is
\be \egw = 4\pi d^2 \frac{c^3}{16 \pi G} \int_{-\infty}^{\infty}\left(\dot{h}_{+}^2 + 
\dot{h}_{\times}^2\right) dt. \ee

We estimate model-dependent upper limits on $\egw$ or $\hrss$ for a given signal 
region as follows:  
\begin{enumerate}[(1)]
\itemsep 1pt
\parskip 0pt
 \item We determine the loudest event in the signal region. 
 \item For a specific simulated signal type, we inject a simulation at a specific $\egw$ and $\hrss
$ in a randomly selected 4\,s interval of the background data and find events 
in that region. We compare the loudest signal region event to the loudest event 
with a cluster centroid time near the known injection time (within 100\,ms for ringdowns and 
within 50\,ms for white noise bursts).  \label{item:inject}
 \item We repeat (\ref{item:inject}) for a range of $\egw$ and $\hrss$ values, and at each value 
we determine the fraction of injections with associated events louder than the loudest signal 
region event. \label{item:repeat}
 \item We repeat (\ref{item:repeat}) using different simulated signal types.  For each
signal type, we estimate the 90\% detection efficiency loudest event upper
limit, $\egwn$ or $\hrssn$, at which 90\% of injection events would be louder than the 
loudest signal region event.
\end{enumerate}

\section{Results and discussion} 
We find no evidence of a GW signal in any of the signal regions analyzed.  
The loudest event of the search occurred at 2009 January 22 05:48:43.2 UTC and was the
only event with a false-alarm rate below our predetermined follow-up
threshold of $1/(3 \times 4808\,\mathrm{s})$ = $\sci{6.9}{-5}$\,Hz as
estimated via extrapolation from the 2000\,s local background
region.   This event cannot be considered a GW candidate because
it was found when only the Hanford 2\,km detector was observing and was
coincident with a strong glitch caused by fluctuations in the
AC power picked up by a magnetometer, and thus is highly likely to be
an instrumental artifact.

We estimate $\hrssn$ and $\egwn$ for each signal region, which depend on detector sensitivities and antenna 
factors, the loudest signal region event, and the simulation waveform 
type.  $\egwn$ upper limits also depend on nominal source distance
$d_{\mathrm{N}}$ and can be scaled to any source distance $d$ via the factor $(d/
d_{\mathrm{N}})^2$.  

\begin{figure*}[!t]
\begin{center}
\includegraphics[angle=0,width=200mm, viewport=110 0 1000 350, clip]{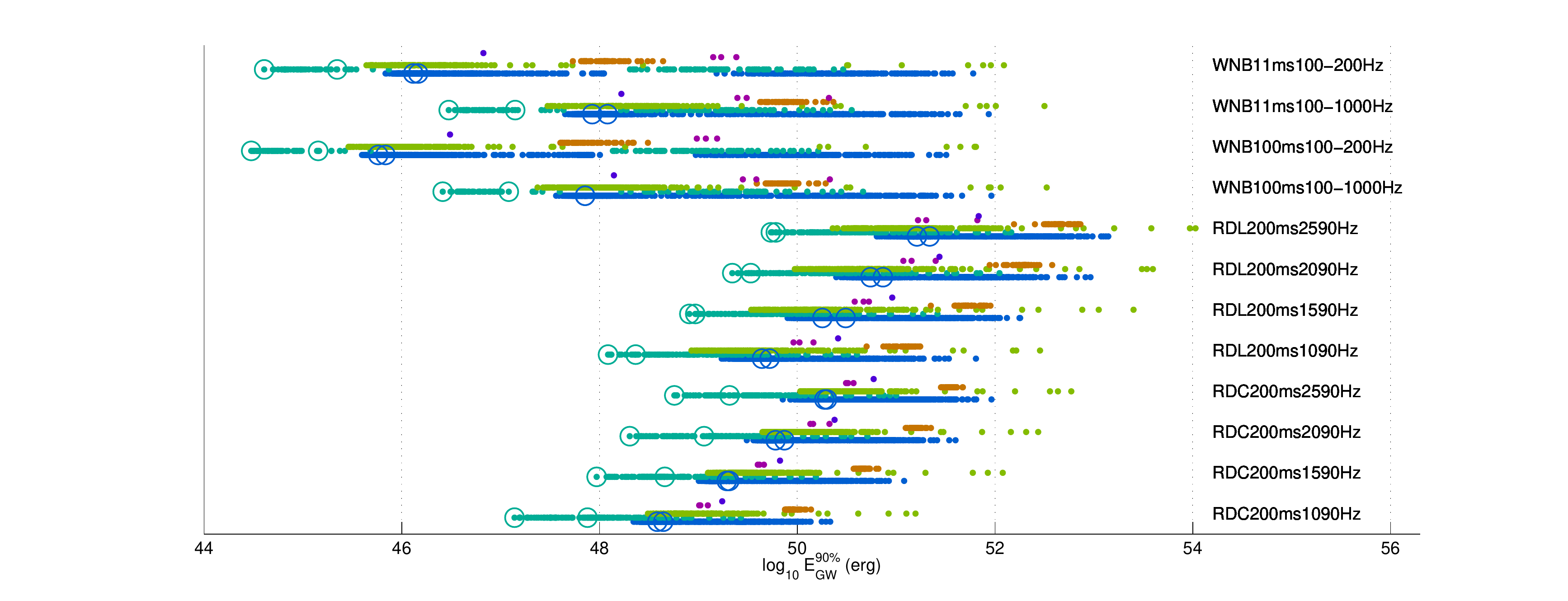}
\caption{ $\egwn$ upper limits for the entire SGR burst sample for
various circularly/linearly polarized ringdowns (RDC/RDL) and white noise
burst (WNB) signals (see Section\,\ref{sec:method}).  For each of twelve waveform types, 
we show six rows of dots marking upper limits for the sources (from top to bottom)  
SGR 1900+14 (violet); SGR 0418+5729 (purple); SGR 1627--41 (orange); 
SGR 1806$-$20 (green); SGR 0501+4516 (teal); and AXP 1E 1547.0-4508 (blue) for that
waveform type.
The limits shown in Table\,\ref{table:bestresults} for SGR 0501+4516 and 
AXP 1E 1547.0-4508 are indicated in the figure by circles.  } \label{fig:egwAll}
\end{center}
\end{figure*}

Figure\,\ref{fig:egwAll} shows $\egw$ upper limits for each of the EM triggers from the six 
magnetar candidates, for each waveform type.  The complete table of upper limits is online\footnote{https://dcc.ligo.org/cgi-bin/DocDB/ShowDocument?docid=25737}.
We spotlight bright bursts from SGR 0501+4516 and AXP 1E
1547.0$-$5408; however it is unknown whether $\eem$ and
$\egw$ are correlated.
Table\,\ref{table:bestresults} presents $\egw$ and $\hrss$ upper limits for three exceptional 
EM triggers; and for a burst from SGR 0501+4516 occurring in the signal region centered at 
2008 August 23 16:31:22 UTC which yielded the lowest limits of the search.  Each 
was analyzed with a network of the LIGO 2\,km and GEO600 
detectors.  The SGR 0501+4516 burst with the largest EM fluence ($\sci{2.21}{-5}$\,erg/cm
$^2$\,\citep{aptekar09}, which corresponds to a 1\,kpc isotropic energy of $\sci{2.7}{39}$\,erg) 
occurred in the signal region centered at 2008 August 24 01:17:58 UTC.  The two candidate 
progenitor bursts for the expanding X-ray rings around AXP 1E 1547.0$-$5408 occurred at 
2009 January 22 6:45:14 UTC and 6:48:04 UTC, with estimated isotropic $\eem$ of 
$10^{44-45}$\,erg\,\citep{tiengo10}.  Table\,\ref{table:bestresults} also gives upper limits on the 
ratio $\gamma \equiv \egwn / \eem$ for the three bursts with $\eem$ estimates.  The $
\gamma$ upper limits for the two ring bursts were estimated using $\eem=10^{45}$\,erg, and 
beat the best previous upper limits on $\gamma$, set for the SGR 1806$-$20 GF\,
\citep{s5y1sgr}, by a factor of a few. 

Superscripts in Table\,\ref{table:bestresults} give uncertainties at 90\% confidence. The first 
is uncertainty in detector amplitude
calibrations. The second is the statistical uncertainty (via the bootstrap method) from using a finite number of 
injected simulations.  Both are added linearly to final $\hrss$ upper limit estimates; corresponding 
uncertainties are added to $\egw$ upper limit estimates.

Our best $\egw$ $f$-mode upper limits are an order of magnitude lower (better) than the best $f$-mode
limits from previous searches, and approach the range of EM energies seen in
SGR GFs for the first time.  The best SGR 0501+4516 $f$-mode limit of
$\sci{1.4}{47}$\,erg (at 1090\,Hz and a nominal distance of 1\,kpc) probes
below the available energy predicted in a fraction of the
parameter space explored in \cite{ioka01} and \cite{corsi11}, the predicted maximum being 
$\sim 10^{48}$--$10^{49}$\,erg.  The best 100--200\,Hz white noise burst limit of
$\sci{3.5}{44}$\,erg is---for the first time---comparable to the $\eem$ seen in ``normal'' GFs.

Improved upper limits and perhaps detection will come in the future via the following routes:  
\begin{enumerate}[(1)]
\itemsep 1pt
\parskip 0pt
\item Additional GFs could push down upper limits on $\gamma \equiv \egwn / \eem$.
\item An analysis which stacks isolated bursts (from e.g. SGR 0501+4516 and AXP 1E 1547.0$-$5408) using the method 
of\,\cite{stackSgr}.  Stacking 100 or more bursts observed with a constant detector sensitivity, 
as in\,\cite{stackSgr}, might yield up to an additional order of magnitude improvement in $\egwn$.  
\item The GW detectors will become more sensitive.  Second generation detectors 
(Advanced LIGO and Advanced Virgo) are expected to begin observing by 2015, promising 
more than two orders of magnitude improvement in $\egw$ sensitivity over the LIGO 2\,km + 
GEO600 network which observed SGR 0501+4516\,\citep{ligoDetector}.   
Recently \cite{levin11} made semi-quantitative predictions on $f$-mode excitations in GFs.
Their predictions are pessimistic that an $f$-mode signal from a GF at 1\,kpc would be detectable in 
the second generation, though they do not consider crustal cracking. 
Third generation detectors could yield two additional orders of magnitude in energy 
sensitivity.  
\end{enumerate}

We look forward to further predictions on GW emission amplitudes from these enigmatic sources.

\begin{table*}
\begin{center}
\caption{GW strain and energy upper limit estimates at 90\% confidence ($\hrssn$ and $\egwn$), for the burst trigger yielding the lowest $\egwn$ upper limits (top left), the brightest SGR 0501+4516 burst (top right), and the two `ring' events from AXP 1E 1547.0$-$5408 (bottom).  Upper limits on the ratio $\gamma \equiv \egwn / \eem$ are given when estimates for $\eem$ are available; for the ring events $\gamma = \egwn / 10^{45}$\,erg. Upper limits were estimated using the circularly and linearly polarized ringdowns (RDC/RDL) and white noise burst (WNB) waveforms (see Section\,\ref{sec:method}). Uncertainties, from detector calibration and using a finite number of injected simulations, are added to the final upper limit estimates.  These are given for the $\hrss$ limits as superscripts, with the first showing detector calibration uncertainty and the second showing statistical uncertainty from finite injected simulations.}

\begin{tabular}{@{\extracolsep{\fill}}lrlr|r|c||rlr|r|c}
 \hline \hline
 & \multicolumn{5}{c}{SGR 0501+4516 Best Limits} & \multicolumn{5}{c}{SGR 0501+4516 Brightest Burst} \\ 
 Simulation type & \multicolumn{3}{c}{$ \hrssn ( 10^{-22}~ \rthz $) }  & $\egwn$ (erg) & $\gamma$  & \multicolumn{3}{c}{$ \hrssn ( 10^{-22} ~ \rthz $) }  & $\egwn$ (erg)  & $\gamma$  \\ 

 \hline 
 WNB 11ms 100-200 Hz   & 7.0 & $^{ +1.0 ~ +0.89}$ &  $= 8.9$ & $\sci{6.8}{44}$ & -  & 13 & $^{ +1.9 ~ +1.3}$ &  $= 16$ & $\sci{2.2}{45}$ & $\sci{8}{5}$ \\
 WNB 100ms 100-200 Hz   & 5.1 & $^{ +0.76 ~ +0.26}$ &  $= 6.1$ & $\sci{3.1}{44}$ & -  & 11 & $^{ +1.6 ~ +0.69}$ &  $= 13$ & $\sci{1.4}{45}$ & $\sci{5}{5}$ \\
 WNB 11ms 100-1000 Hz   & 13 & $^{ +3.6 ~ +0.62}$ &  $= 17$ & $\sci{3.5}{46}$ & -  & 25 & $^{ +7.3 ~ +1.6}$ &  $= 34$ & $\sci{1.4}{47}$ & $\sci{5}{7}$ \\
 WNB 100ms 100-1000 Hz   & 13 & $^{ +3.7 ~ +0.56}$ &  $= 17$ & $\sci{3.2}{46}$ & -  & 26 & $^{ +7.3 ~ +1.4}$ &  $= 34$ & $\sci{1.2}{47}$ & $\sci{4}{7}$ \\
 RDC 200ms 1090 Hz   & 15 & $^{ +2.4 ~ +1.3}$ &  $= 18$ & $\sci{1.4}{47}$ & -  & 35 & $^{ +5.8 ~ +1.7}$ &  $= 42$ & $\sci{7.6}{47}$ & $\sci{3}{8}$ \\
 RDC 200ms 1590 Hz   & 30 & $^{ +4.9 ~ +2.1}$ &  $= 37$ & $\sci{1.2}{48}$ & -  & 59 & $^{ +9.7 ~ +2.9}$ &  $= 71$ & $\sci{4.6}{48}$ & $\sci{2}{9}$ \\
 RDC 200ms 2090 Hz   & 32 & $^{ +5.3 ~ +1.6}$ &  $= 39$ & $\sci{2.4}{48}$ & -  & 69 & $^{ +11 ~ +4.7}$ &  $= 85$ & $\sci{1.1}{49}$ & $\sci{4}{9}$ \\
 RDC 200ms 2590 Hz   & 40 & $^{ +6.7 ~ +2.9}$ &  $= 50$ & $\sci{6.1}{48}$ & -  & 77 & $^{ +13 ~ +5.2}$ &  $= 96$ & $\sci{2.1}{49}$ & $\sci{8}{9}$ \\
 RDL 200ms 1090 Hz   & 40 & $^{ +6.6 ~ +8.1}$ &  $= 54$ & $\sci{1.3}{48}$ & -  & 58 & $^{ +9.6 ~ +5.0}$ &  $= 73$ & $\sci{2.3}{48}$ & $\sci{9}{8}$ \\
 RDL 200ms 1590 Hz   & 86 & $^{ +14 ~ +9.8}$ &  $= 110$ & $\sci{1.1}{49}$ & -  & 80 & $^{ +13 ~ +6.5}$ &  $= 100$ & $\sci{9.3}{48}$ & $\sci{3}{9}$ \\
 RDL 200ms 2090 Hz   & 96 & $^{ +16 ~ +20}$ &  $= 130$ & $\sci{2.7}{49}$ & -  & 110 & $^{ +19 ~ +12}$ &  $= 140$ & $\sci{3.4}{49}$ & $\sci{1}{10}$ \\
 RDL 200ms 2590 Hz   & 110 & $^{ +19 ~ +17}$ &  $= 150$ & $\sci{5.5}{49}$ & -  & 120 & $^{ +21 ~ +10}$ &  $= 160$ & $\sci{6.1}{49}$ & $\sci{2}{10}$ \\

 \hline 
\\
 & \multicolumn{5}{c}{AXP 1E 1547.0$-$5408 2009 Jan 22 6:45:14 UTC} & \multicolumn{5}{c}{AXP 1E 1547.0$-$5408 2009 Jan 22 6:48:04 UTC} \\ 
 \hline 
 WNB 11ms 100-200 Hz   & 7.9 & $^{ +1.2 ~ +1.3}$ &  $= 10$ & $\sci{1.5}{46}$ & 10  & 7.6 & $^{ +1.1 ~ +1.0}$ &  $= 9.8$ & $\sci{1.3}{46}$ & 10 \\
 WNB 100ms 100-200 Hz   & 5.3 & $^{ +0.80 ~ +0.41}$ &  $= 6.6$ & $\sci{5.8}{45}$ & 6   & 5.8 & $^{ +0.86 ~ +0.47}$ &  $= 7.1$ & $\sci{6.8}{45}$ & 7  \\
 WNB 11ms 100-1000 Hz   & 16 & $^{ +4.5 ~ +1.0}$ &  $= 21$ & $\sci{8.4}{47}$ & $\sci{8}{2}$  & 18 & $^{ +5.2 ~ +0.91}$ &  $= 24$ & $\sci{1.2}{48}$ & $\sci{1}{3}$ \\
 WNB 100ms 100-1000 Hz   & 15 & $^{ +4.5 ~ +0.73}$ &  $= 21$ & $\sci{7.1}{47}$ & $\sci{7}{2}$  & 16 & $^{ +4.5 ~ +0.80}$ &  $= 21$ & $\sci{7.2}{47}$ & $\sci{7}{2}$ \\
 RDC 200ms 1090 Hz   & 19 & $^{ +3.2 ~ +1.4}$ &  $= 24$ & $\sci{3.8}{48}$ & $\sci{4}{3}$  & 21 & $^{ +3.5 ~ +1.0}$ &  $= 25$ & $\sci{4.4}{48}$ & $\sci{4}{3}$ \\
 RDC 200ms 1590 Hz   & 30 & $^{ +4.9 ~ +2.5}$ &  $= 37$ & $\sci{1.9}{49}$ & $\sci{2}{4}$  & 31 & $^{ +5.1 ~ +1.6}$ &  $= 37$ & $\sci{2.1}{49}$ & $\sci{2}{4}$ \\
 RDC 200ms 2090 Hz   & 39 & $^{ +6.6 ~ +2.8}$ &  $= 49$ & $\sci{6.0}{49}$ & $\sci{6}{4}$  & 44 & $^{ +7.4 ~ +3.8}$ &  $= 56$ & $\sci{7.4}{49}$ & $\sci{7}{4}$ \\
 RDC 200ms 2590 Hz   & 57 & $^{ +9.4 ~ +3.9}$ &  $= 70$ & $\sci{1.9}{50}$ & $\sci{2}{5}$  & 60 & $^{ +1.00 ~ +3.3}$ &  $= 73$ & $\sci{2.0}{50}$ & $\sci{2}{5}$ \\
 RDL 200ms 1090 Hz   & 60 & $^{ +1.00 ~ +9.9}$ &  $= 80$ & $\sci{4.4}{49}$ & $\sci{4}{4}$  & 66 & $^{ +11 ~ +10}$ &  $= 87$ & $\sci{5.3}{49}$ & $\sci{5}{4}$ \\
 RDL 200ms 1590 Hz   & 84 & $^{ +14 ~ +13}$ &  $= 110$ & $\sci{1.8}{50}$ & $\sci{2}{5}$  & 110 & $^{ +18 ~ +22}$ &  $= 150$ & $\sci{3.1}{50}$ & $\sci{3}{5}$ \\
 RDL 200ms 2090 Hz   & 110 & $^{ +19 ~ +16}$ &  $= 150$ & $\sci{5.5}{50}$ & $\sci{6}{5}$  & 130 & $^{ +22 ~ +18}$ &  $= 170$ & $\sci{7.4}{50}$ & $\sci{7}{5}$ \\
 RDL 200ms 2590 Hz   & 150 & $^{ +25 ~ +33}$ &  $= 210$ & $\sci{1.6}{51}$ & $\sci{2}{6}$  & 180 & $^{ +30 ~ +29}$ &  $= 240$ & $\sci{2.2}{51}$ & $\sci{2}{6}$ \\
 \hline \label{table:bestresults} 
 \end{tabular} 
 \end{center} 
 \end{table*}

\acknowledgments
The authors gratefully acknowledge the support of the United States
National Science Foundation for the construction and operation of the
LIGO Laboratory, the Science and Technology Facilities Council of the
United Kingdom, the Max-Planck-Society, and the State of
Niedersachsen/Germany for support of the construction and operation of
the GEO600 detector, and the Italian Istituto Nazionale di Fisica
Nucleare and the French Centre National de la Recherche Scientifique
for the construction and operation of the Virgo detector. The authors
also gratefully acknowledge the support of the research by these
agencies and by the Australian Research Council, the Council of
Scientific and Industrial Research of India, the Istituto Nazionale di
Fisica Nucleare of Italy, the Spanish Ministerio de Educaci\'on y
Ciencia, the Conselleria d'Economia Hisenda i Innovaci\'o of the
Govern de les Illes Balears, the Foundation for Fundamental Research
on Matter supported by the Netherlands Organisation for Scientific Research, 
the Polish Ministry of Science and Higher Education, the FOCUS
Programme of Foundation for Polish Science,
the Royal Society, the Scottish Funding Council, the
Scottish Universities Physics Alliance, The National Aeronautics and
Space Administration, the Carnegie Trust, the Leverhulme Trust, the
David and Lucile Packard Foundation, the Research Corporation, and
the Alfred P. Sloan Foundation.
The Konus-Wind experiment is supported by a Russian Space Agency
contract and RFBR grant 09-02-00166a.
AJvdH is supported by the NASA Postdoctoral Program. 
KH acknowledges IPN support under the following grants:
JPL Y503559 (Odyssey); NASA NNG06GH00G, NASA NNX07AM42G, and NASA
NNX08AC89G (INTEGRAL); NASA NNG06GI896, NASA NNX07AJ65G, and NASA
NNX08AN23G (Swift); NASA NNX07AR71G (MESSENGER); NASA NNX06AI36G, 
NASA NNX08AB84G, NASA NNX08AZ85G (Suzaku); and NASA NNX09AU03G (Fermi). 
YK acknowledges EU FP6 Project MTKD-CT-2006-042722.  
CK acknowledges support from NASA grant NNH07ZDA001-GLAST.  
This Letter is LIGO-\ligodoc{}.

\bibliographystyle{apj}

\end{document}